\documentclass[apjl]{surelythiscantbenecessaryapj}
\usepackage[utf8]{inputenc}
\usepackage{natbib}
\usepackage[usenames]{color}
\usepackage{graphicx}
\usepackage{multirow}
\usepackage{makecell}
\usepackage{microtype}
\usepackage{upgreek}
\usepackage{hyperref}

\newcommand{\dg}     {\mbox{$^{\circ}$}}
\newcommand{\lt}     {\mbox{$<$}}

\newcommand{\narrow} {\fontfamily{phv}\fontseries{mc}\selectfont}
\newcommand{\dfn}[1] {{\bf #1}}

\newcommand{\gtsim}  {\lower.5ex\hbox{$\; \buildrel > \over \sim \;$}}
\newcommand{\ltsim}  {\lower.5ex\hbox{$\; \buildrel < \over \sim \;$}}

\definecolor{olive}{rgb}{0.5, 0.5, 0.0}

\newcommand{\eventA}{ACT-T J181515-492746}
\newcommand{\eventB}{ACT-T J070038-111436}
\newcommand{\eventC}{ACT-T J200758+160954}

\begin{document}

\title{The Atacama Cosmology Telescope: Detection of mm-wave transient sources\vspace*{5mm}}

\author{Sigurd~Naess\footnote{snaess@flatironinstitute.org}}
\affiliation{Center for Computational Astrophysics, Flatiron Institute, New York, NY, USA 10010}
\author{Nick~Battaglia}
\affiliation{Department of Astronomy, Cornell University, Ithaca, NY 14853, USA}
\author{~J.Richard~Bond}
\affiliation{Canadian Institute for Theoretical Astrophysics, 60 St. George Street, University of Toronto, Toronto, ON, M5S 3H8, Canada}
\author{Erminia~Calabrese}
\affiliation{School of Physics and Astronomy, Cardiff University, The Parade, Cardiff, Wales, UK CF24 3AA}
\author{Steve~K.~Choi}
\affiliation{Department of Physics, Cornell University, Ithaca, NY 14853, USA}
\affiliation{Department of Astronomy, Cornell University, Ithaca, NY 14853, USA}
\author{Nicholas~F.~Cothard}
\affiliation{Department of Physics, Cornell University, Ithaca, NY 14853, USA}
\affiliation{Department of Applied and Engineering Physics, Cornell University, Ithaca, NY 14853, USA}
\author{Mark~Devlin}
\affiliation{Department of Physics and Astronomy, University of Pennsylvania, 209 South 33rd Street, Philadelphia, PA, USA 19104}
\author{Cody~J.~Duell}
\affiliation{Department of Physics, Cornell University, Ithaca, NY 14853, USA}
\author{Adriaan~J.~Duivenvoorden}
\affiliation{Joseph Henry Laboratories of Physics, Jadwin Hall, Princeton University, Princeton, NJ, USA 08544}
\author{Jo~Dunkley}
\affiliation{Department of Astrophysical Sciences, Peyton Hall, Princeton University, Princeton, NJ, USA 08544}
\author{Rolando~D\"unner}
\affiliation{Instituto de Astrof\'isica and Centro de Astro-Ingenier\'ia, Facultad de F\'isica, Pontificia Universidad Cat\'olica de Chile, Av. Vicu\~na Mackenna 4860, 7820436, Macul, Santiago, Chile}
\author{Patricio~A.~Gallardo}
\affiliation{Department of Physics, Cornell University, Ithaca, NY 14853, USA}
\author{Megan~Gralla}
\affiliation{Department of Astronomy/Steward Observatory, University of Arizona, 933 N. Cherry Ave., Tucson, AZ 85721, USA}
\author{Yilun~Guan}
\affiliation{Department of Physics and Astronomy, University of Pittsburgh, Pittsburgh, PA, USA 15260}
\author{Mark~Halpern}
\affiliation{Department of Physics and Astronomy, University of British Columbia, Vancouver, BC, Canada V6T 1Z4}
\author{J.~Colin~Hill}
\affiliation{Department of Physics, Columbia University, New York, NY, USA 10027}
\affiliation{Center for Computational Astrophysics, Flatiron Institute, New York, NY, USA 10010}
\author{Matt~Hilton}
\affiliation{Astrophysics Research Centre, University of KwaZulu-Natal, Westville Campus, Durban 4041, South Africa}
\author{Kevin~M.~Huffenberger}
\affiliation{Department of Physics, Florida State University, Tallahassee, FL 32306, USA}
\author{Brian~J.~Koopman}
\affiliation{Department of Physics, Yale University, New Haven, CT 06520, USA}
\author{Arthur~B.~Kosowsky}
\affiliation{Department of Physics and Astronomy, University of Pittsburgh, Pittsburgh, PA, USA 15260}
\author{Mathew~S.~Madhavacheril}
\affiliation{Perimeter Institute for Theoretical Physics, Waterloo, ON, Canada N2L 2Y5}
\author{Jeff~McMahon}
\affiliation{Kavli Institute for Cosmological Physics, University of Chicago, Chicago, IL 60637, USA}
\affiliation{Department of Astronomy and Astrophysics, University of Chicago, Chicago, IL 60637, USA}
\affiliation{Department of Physics, University of Chicago, Chicago, IL 60637, USA}
\affiliation{Enrico Fermi Institute, University of Chicago, Chicago, IL 60637, USA}
\author{Federico Nati}
\affiliation{Department of Physics, University of Milano-Bicocca, Piazza della Scienza 3, 20126 Milano (MI), Italy}
\author{Michael~D.~Niemack}
\affiliation{Department of Physics, Cornell University, Ithaca, NY 14853, USA}
\affiliation{Department of Astronomy, Cornell University, Ithaca, NY 14853, USA}
\author{Lyman~Page}
\affiliation{Joseph Henry Laboratories of Physics, Jadwin Hall, Princeton University, Princeton, NJ, USA 08544}
\author{Bruce~Partridge}
\affiliation{Department of Physics and Astronomy, Haverford College, Haverford, PA, USA 19041}
\author{Maria~Salatino}
\affiliation{Physics Department, Stanford University Kavli Institute for Particle Astrophysics and Cosmology (KIPAC) Stanford, California CA}
\author{Neelima~Sehgal}
\affiliation{Physics and Astronomy Department, Stony Brook University, Stony Brook, NY 11794}
\author{David~Spergel}
\affiliation{Center for Computational Astrophysics, Flatiron Institute, New York, NY, USA 10010}
\affiliation{Department of Astrophysical Sciences, Peyton Hall, Princeton University, Princeton, NJ, USA 08544}
\author{Suzanne~Staggs}
\affiliation{Joseph Henry Laboratories of Physics, Jadwin Hall, Princeton University, Princeton, NJ, USA 08544}
\author{Edward~J.~Wollack}
\affiliation{NASA/Goddard Space Flight Center, Greenbelt, MD, USA 20771}
\author{Zhilei Xu}
\affiliation{Department of Physics and Astronomy, University of Pennsylvania, 209 South 33rd Street, Philadelphia, PA, USA 19104}
\affiliation{MIT Kavli Institute, Massachusetts Institute of Technology, Cambridge, MA, USA}

\begin{abstract}
We report on the serendipitous discovery of three transient mm-wave sources using data from the Atacama
Cosmology Telescope.
The first, detected at RA = 273.8138, dec = -49.4628 at ${\sim}50\sigma$ total, brightened from less than 5 mJy to at least 1100 mJy at 150 GHz with an unknown rise time shorter than thirteen days, during which the increase from 250 mJy to 1100 mJy took only 8 minutes. Maximum flux was observed on 2019-11-8. The source's spectral index in flux between 90 and 150 GHz was positive,  $\alpha = 1.5\pm0.2$. The second, detected at RA = 105.1584, dec = -11.2434 at ${\sim}20\sigma$ total, brightened from less than 20 mJy to at least 300 mJy at 150 GHz with an unknown rise time shorter than eight days. Maximum flux was observed on 2019-12-15. Its spectral index was also positive, $\alpha = 1.8\pm0.2$. The third, detected at RA = 301.9952, dec = 16.1652 at ${\sim}40\sigma$ total, brightened from less than 8 mJy to at least 300 mJy at 150 GHz over a day or less but decayed over a few days. Maximum flux was observed on 2018-9-11. Its spectrum was approximately flat, with a spectral index of $\alpha = -0.2\pm0.1$. None of the sources were polarized to the limits of these measurements. The two rising-spectrum sources are coincident in position with M and K stars, while the third is coincident with a G star.
\end{abstract}

\section{Introduction}
\label{sec:intro}


New large-area surveys by sensitive, arcminute-resolution CMB telescopes are making systematic searches for mm-wave transient sources possible.
Although transient phenomena in the mm-wavelength range are relatively unexplored, a variety of sources have been observed.

In an untargeted survey by the South Pole Telescope,
\citet{whitehorn/etal:2016} reported a candidate transient at $2.6\sigma$ significance with a duration of 1 week.  It rose to a peak flux of $16.5\pm 2.4$ mJy at 150~GHz and had a rising spectrum 
and linear polarization.  The possible source's identity is uncertain, but some properties agree with models of GRB afterglows.  Previously, \cite{2004PASJ...56L...1K} observed GRB 030329's afterglow at 90 GHz and measured it at $\sim 65$ mJy (at $z=0.17$, nearby for a GRB). Additionally \citet{laskar-grb-2019} observed  at 97.5 GHz a polarized GRB afterglow associated with a reverse shock. Generally, millimeter afterglows from GRBs last a few days to about a week \citep{2012A&A...538A..44D}.

Other types of transient events are also seen.  The relatively nearby (at 17 Mpc) tidal disruption event IGR J12580+0134  was serendipitously captured in the six bands of the Planck High-Frequency Instrument, from 100--857 GHz \citep{2016MNRAS.461.3375Y}; the flux in each of these bands exceeded 400 mJy and showed a flat or slightly declining spectrum.  The event lasted less than a year.  This event may have been a sub-stellar object torn apart by a supermassive black hole \citep{Lei_2016}.
\citet{ho/etal:2019} presented millimeter wave measurements of the exceptional transient AT2018cow.  They argued that it represents a new class of extragalactic transients sourced by X-ray emission shocking a dense medium.  This source maintained its mm-wave brightness for several tens of days. The spectrum rose at frequencies lower than roughly 100 GHz and fell for higher frequencies.  Other observed mm-wave transients include stellar flares from T Tauri stars and other young stellar objects, found in observations that targeted star-forming regions.  The flares are attributed to magnetic reconnection events \citep{Bower_2003,Massi_2006,Mairs_2019}.  The spectra of these events can rise or fall across the millimeter (and change with time).

Synchrotron emission is an important mechanism in many of these phenomena.  If we characterize an object's flux density as $S_\nu\propto\nu^\alpha$, the synchrotron spectrum at low frequencies rises ($\alpha>0$) due to self-absorption in the optically thick regime \citep{1986rpa..book.....R}.  At higher frequencies, the medium is optically thin, and the spectrum falls ($\alpha < 0$), following the distribution of energetic electrons.
The spectrum tells us about the condition of the emitting region.  The most common variable sources between 90 and 150 GHz, active galactic nuclei (AGN), have $\alpha\ltsim 0$. Thus transients with $\alpha>0$ between these frequencies are especially notable.

We report here on three serendipitous, high-significance discoveries of transient sources in maps from the Atacama Cosmology Telescope (ACT).  The events are relatively bright, represent large fractional increases in luminosity, and appear to be unpolarized.
	One has rapid, minute-timescale brightening, and two have rising spectra.  These transients appear to be associated with stars.

\section{Observations }
\label{sec:obs}
\subsection{ACT}
The ACT experiment \citep{thornton/etal:2016} is a 6-meter telescope on Cerro Toco in the northern Chilean Andes. The third-generation receiver \citep[AdvACT,][]{henderson/etal:2016,choi/etal:2018,crowley/etal:2018} houses three separate arrays of feedhorn-coupled, dichroic, dual-polarization, transition-edge-sensor (TES) bolometers from the United States
National Institute of Standards and Technology (NIST), with each array occupying an individual optics tube. The arrays are cooled to 0.1 K and detect radiation in broad bands centered on 98, 150, and 225 GHz \citep{choi/etal:2020}. We label these bands f090, f150, and f220 respectively. The bolometers are read out with time-domain multiplexing electronics \citep[chapter 2]{matthew-thesis} and stored in ``time-ordered data'' files (TODs), each of roughly 10-minute duration. Detector arrays PA5 and PA6 contain dichroic detectors which record intensity and linear polarization in the f090 and f150 bands simultaneously in each of 429 feeds. Array PA4 is similar but for the f150 and f220 bands in each of 503 feeds. Each optics tube images an instantaneous field of view with $\sim 1^\circ$ diameter on the sky. The average beam full-width-at-half-maximum for each band is 2.05/1.40/0.98 arcminutes at f090/f150/f220 respectively.

The observing strategy scans the telescope in azimuth with a peak-to-peak amplitude near $60^\circ$ and a one way scan time of roughly 40 sec \citep{debernardis/etal:2016,choi/etal:2020}. An individual scan covers a
stripe of the sky the length of the azimuth throw and the width of the detector arrays, but due to the rotation
of the sky the covered area drifts by $15^\circ$ per hour in RA, allowing the same scanning motion to cover
large areas of sky. When scanning in the East, a point on the sky is first observed
when the bottom-most detectors in the lower two arrays (PA4 and PA5)
sweep across it.  These are gradually followed by detectors higher in the focal plane as the source rises.
After $\sim 6$ minutes it rises above these arrays, and about 3 minutes later enters the bottom
of the upper array, PA6, where the process repeats. When scanning in the West the setting source crosses the arrays in reverse, from top to bottom.
All in all, a point on the sky takes about 15 minutes to traverse the ACT focal plane.

Since 2016 ACT has been surveying $\sim$40\% of the sky with a variable but roughly weekly cadence, resulting in a
set of 200 megapixel sky maps \citep[e.g.][]{naess/etal:2020}. ACT calibrates the data in several
ways, including by cross-correlation to quarter-degree-scale CMB fluctuations in maps from the Planck satellite.
ACT observes both during the night and day. At night, the telescope has a simpler optical response and the processing of nighttime data is more mature. All three transient sources were observed at night.

Based on comparing our mapping solution to the positions of known sources, pointing accuracy for the analysis presented here is approximately 3 arcsec (0.05 arcmin) independently in Dec and RA/cos(Dec). This translates into effective 1$\sigma$, 2$\sigma$, and 3$\sigma$ radii of
5.8, 8.3 and 10.9 arcsec respectively for a $\chi$ distribution with two degrees of freedom.

\subsection{Discovery}
The events presented here were not found in a search optimized for transients, but were instead
serendipitously discovered during investigation of candidate events in a search for the hypothetical
Planet 9 \citep{p9_hypothesis}, which will be the subject of an upcoming paper \citep[in preparation]{act-planet9}.
This search covers roughly
18\,000 square degrees of sky using almost all ACT data from 2008 to 2019. However, the shift-and-stack
algorithm used for the planet search is far from optimal for detecting transients, so the events
presented here were only found due to their high brightness. We expect a forthcoming search optimized
specifically for transients to have a higher yield.

\subsection{Characterization}
We perform a maximum likelihood fit of the position and a per-detector-array, per-frequency
flux for each of our three detections. This fit was performed directly in the time-ordered data,
using the same noise model used for our normal map-making \citep[e.g.,][]{aiola/etal:2020},
and therefore takes into account
both the temporal and inter-detector correlation structure of the atmospheric and instrumental
noise. Due to ACT's broad scans, each TOD hits multiple bright point sources with known positions,
in addition to the transient. We improve the absolute position accuracy by including these in the fit.

\begin{figure*}[htbp]
\begin{center}
\includegraphics[width=0.9\textwidth]{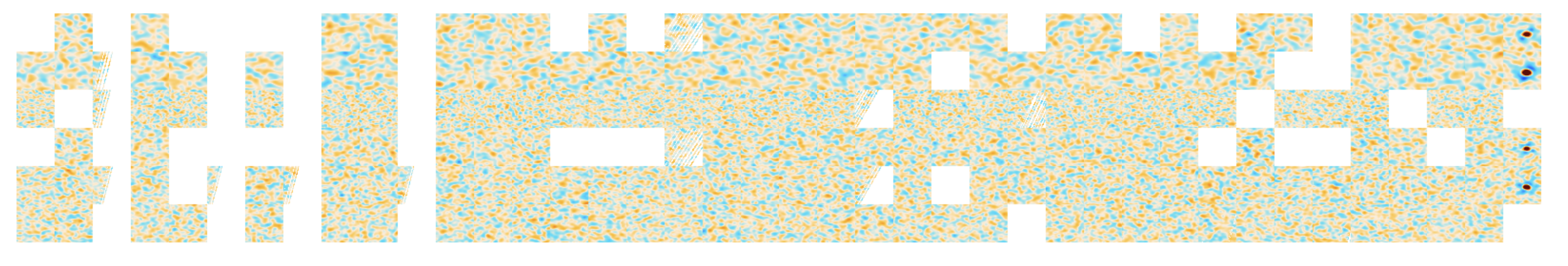}
	\caption{Diagnostic plot for Event 1.  The rows show a series of $0.5^{\circ}\times0.5^{\circ}$ filtered maps corresponding to PA4 f150, PA5 f150, PA6 f150, PA4 f220, PA5 f090 and PA6 f090 from the bottom to the top, while the columns correspond to individual 3-day chunks of data for the 2017-2019 observing seasons, with those that don't have exposure near this object omitted (so there are implicit, variable-length gaps between each column). The maps are in units of signal-to-noise ratio, spanning from -8 (blue) to +8 (red). Time increases from left to right. It is clear that this region is quiet until the last map where a strong point source suddenly appears in all arrays that hit it. The columns are not equi-spaced, and there is a gap of 12 days between the preceding thumbnail maps and the ones in which the source is detected. The source is evident in the PA4 time line but the data do not pass our data quality cuts. The time for the last map spans from 2019-11-7 00:00:00 to 2019-11-10 00:00:00. Unfortunately, there are no observations of this region in 2019 after those shown in this figure. There is a several month long gap before any 2020 coverage of this region, and the 2020 data is not yet ready for analysis.
}
\label{fig:thumbnails}
\end{center}
\end{figure*}

\begin{figure}[htbp]
	\begin{center}
	\includegraphics[width=\columnwidth]{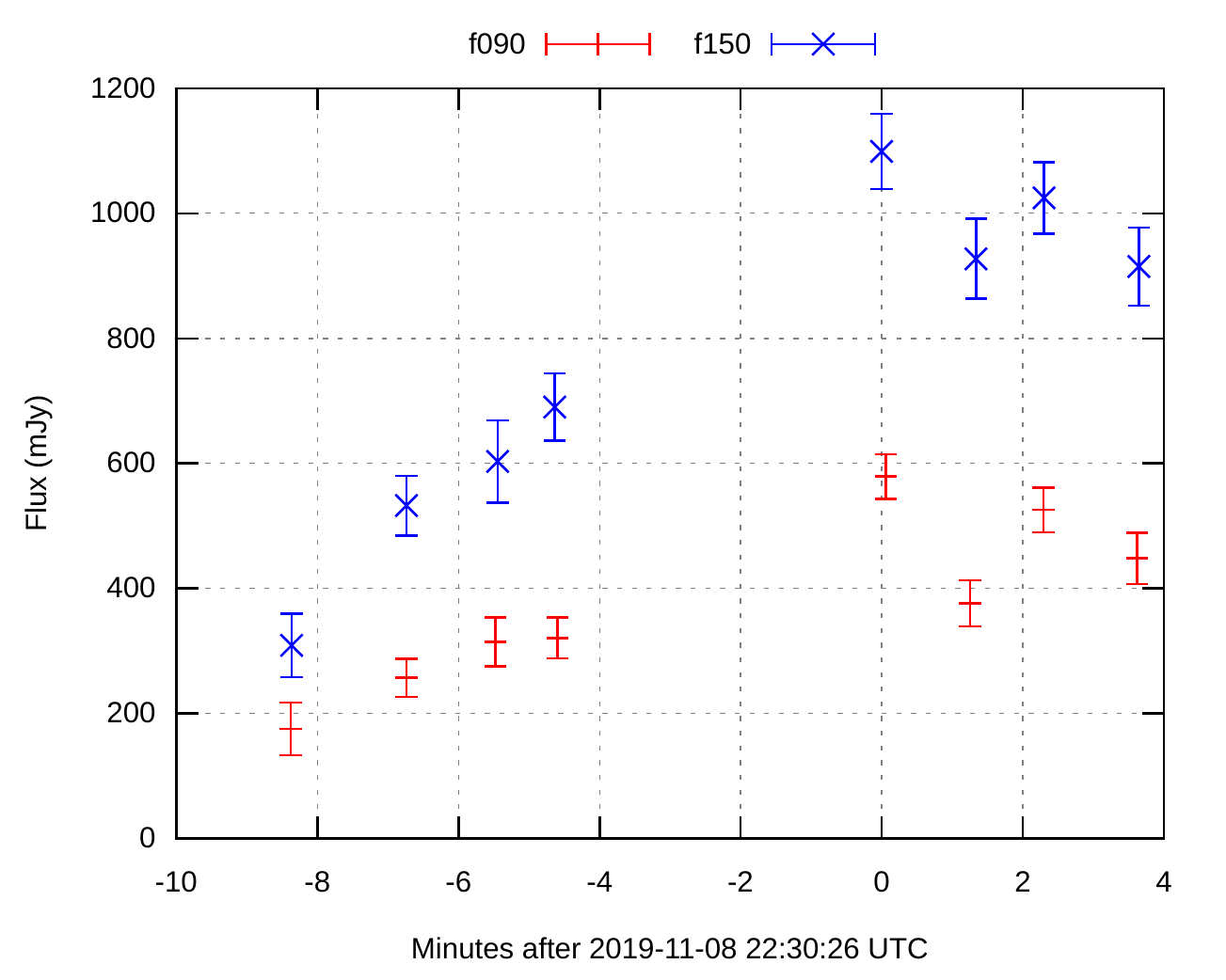}
		\caption{High-resolution light curve for Event~1 \eventA{} made by subdividing the ACT detector arrays
		into 4 sub-sections that hit the source at slightly different times. We observe a rapid rise from
		about 180 to 600 mJy at f090 and from 600 to 1100 mJy at f150 over the course of 8 minutes followed by what might
		be the beginning of a slower decay. We do not detect anything at this location before this, even
		when averaging four years of data. The gap between -4 and 0 minutes on the x-axis corresponds to
		the separation between two of our detector arrays. Unfortunately there are no observations of this
		region in 2019 after this. No f220 data survived our data cuts during the event.}
	\label{fig:e1-lightcurve}
	\end{center}
\end{figure}

\section{The events}
\label{sec:events}

\begin{table*}[hpt]
	\begin{center}
	\caption{Overview of the events}
	\label{tab:overview}
	\begin{tabular}{clrrr}
		&       &  Event 1 & Event 2 & Event 3 \\
		Name &       &  \scriptsize\narrow \eventA{} & \scriptsize\narrow \eventB{} &  \scriptsize\narrow \eventC{} \\
		\hline
		\multirow{4}{*}{\rotatebox[origin=c]{0}{\makecell{Coords\\\dg}}}
		&RA     &            273.8138 &            105.1584 &            301.9952 \\
		&Dec    &            -49.4628 &            -11.2434 &             16.1652 \\
		&l      &             15.4692 &            223.9866 &             56.2617 \\
		&b      &            -14.8090 &             -3.0893 &             -8.8159 \\
		\hline
		\multicolumn{2}{l}{Pos. acc. "} &    3 &  3 &  3 \\
		\hline
		\multirow{3}{*}{\rotatebox[origin=c]{0}{\makecell{Time}}}
		& Peak  & 2019-11-8 22:30 & 2019-12-15 02:22 & 2018-9-11 23:36 \\
		& Rise  & 8 min & $<8$ days & $<1$ day \\
		& Fall  & $\gg 4$ min & $\gg 10$ min        & $\approx$ 2 weeks \\
		\hline
		\multirow{3}{*}{\rotatebox[origin=c]{0}{\makecell{Mean flux\\mJy}}}
		& f090  &           5.9 $\pm$ 4.8 &       -0.7 $\pm$ 5.8 &      10.3 $\pm$ 2.9 \\
		& f150  &          -2.4 $\pm$ 3.3 &        7.7 $\pm$ 6.5 &       3.6 $\pm$ 2.2 \\
		& f220  &          -1.0 $\pm$ 4.0 &       -5.2 $\pm$13.2 &       0.4 $\pm$ 4.5 \\
		\hline
		\multirow{3}{*}{\rotatebox[origin=c]{0}{\makecell{Peak flux\\mJy}}}
		& f090  &             579 $\pm$36 &          143 $\pm$13 &         340 $\pm$10 \\
		& f150  &            1099 $\pm$60 &          304 $\pm$18 &         307 $\pm$14 \\
		& f220  &                  -- &               -- &         315 $\pm$54 \\
		\hline
		\multirow{3}{*}{\rotatebox[origin=c]{0}{\makecell{Min factor\\50\%}}}
		& f090  &                  88 &               39 &                33 \\
		& f150  &                 719 &               35 &                82 \\
		& f220  &                  -- &               -- &                98 \\
		\hline
		\multirow{3}{*}{\rotatebox[origin=c]{0}{\makecell{Min factor\\98\%}}}
		& f090  &                  36 &               11 &                21 \\
		& f150  &                 181 &               14 &                38 \\
		& f220  &                  -- &               -- &                27 \\
		\hline
		\multirow{3}{*}{\rotatebox[origin=c]{0}{\makecell{Pol limit\\mJy}}}
		& f090  &     \lt 76.8 (15\%) &     \lt 74.2 (41\%) &  \lt 56.7 (18\%) \\
		& f150  &     \lt 57.1 (10\%) &     \lt 63.0 (31\%) &  \lt 95.4 (24\%) \\
		& f220  &                  -- &                  -- &  \lt 57.9 (74\%) \\
		\hline
		\multicolumn{2}{l}{Spec.ind. ($\alpha$)} & 1.5 $\pm$ 0.2 & 1.8 $\pm$ 0.2 & -0.2 $\pm$ 0.4 \\
		\hline
		\multirow{4}{*}{Assoc}
		& Name    & \scriptsize\narrow 2MASS J18151564-4927472 & \scriptsize\narrow HD 52385 & \scriptsize\narrow HD 191179 \\
		& Sep. "  & 3.5                              & 10.7     & 6.4     \\
		& Chance  & $3.7\times 10^{-4}$              & $8.9\times 10^{-5}$ & $2.8\times 10^{-5}$ \\
		& Dist pc & 62.0 $\pm$ 0.6                          & 403 $\pm$ 4  & 219 $\pm$ 1 \\ 
		\hline
		\multirow{3}{*}{\rotatebox[origin=c]{0}{\makecell{$\nu L_\nu$\\\tiny$10^{22}$W}}}
		& f090 &    2.61 $\pm$ 0.16 &  27.20 $\pm$ 2.48 &  19.14 $\pm$ 0.59 \\
		& f150 &    7.59 $\pm$ 0.42 &  88.73 $\pm$ 5.13 &  26.38 $\pm$ 1.18 \\
		& f220 &             -- &            -- &  39.73 $\pm$ 6.81 \\
		\hline
		\multicolumn{2}{l}{Notes} &
		\makecell*[{{p{27mm}}}]{\scriptsize Rapid rise from near zero to peak in 8 minutes.
		Blue spectrum.} &
		\makecell*[{{p{27mm}}}]{\scriptsize Not observed during rise and fall. No detectable evolution over 15 min.
		Blue spectrum. Position offset is a bit big.} &
		\makecell*[{{p{27mm}}}]{\scriptsize Flat spectrum. Slow, non-monotonic fall from peak over 2 weeks.} \\
		\hline
	\end{tabular}
	\end{center}
	\tablecomments{\small
		The ``peak'' time represents when ACT observed the
		highest flux from the source, in UTC. Due to sparse coverage this might not coincide with the
		actual peak of the light curve. The rise/fall times are the time it takes the source
		to rise/fall from below detectability to the peak. Due to sparse coverage some of these are
		only upper limits, and some are completely unconstrained because the source was never observed
		after this. The coordinates are J2000 and have an approximate accuracy of 3 arcseconds ($1\sigma$ error), with
		comparable contributions from statistical and systematic uncertainty. The ``mean'' fluxes
		reported are those from forced photometry at the source position on maps using all data from
		the 2016--2019 observing seasons, which includes the event itself as a small subset.
		The ``min factor'' gives the 50\% and 98\% lower bound on how many times brighter the source
		got during the peak compared to its normal state.
		We do not detect polarization for any of the events, from which we derive the
		98\% upper limits in flux and polarization fraction given in the row ``pol limit''.
		The spectral index was measured using f090 and f150 only. Distances come from Gaia DR 2.
		``Chance'' is the probability that an object as bright as the given candidate would be
		that close to the event by chance.
		$\nu L_\nu$ gives the characteristic luminosity of the event in units of $10^{22}$W,
		assuming isotropy and that the event is at the distance given by the tentative association.
		Luminosity error bars do not include distance uncertainty.
	}
\end{table*}

\subsection {Event 1: \eventA{}}
Event 1 was observed at RA$=273.8162^\circ \pm 0.0013^\circ$ and Dec$=-49.4628^\circ \pm 0.00083^\circ$
during one TOD in PA5 and one in PA6 between 22:22 and 22:34 UTC on 2019-11-8. The galactic coordinates
are $l=344.53^\circ$ and $b=-14.81^{\circ}$ consistent with being in the galactic plane.
The two arrays measured
a large and unexpected difference in the event's flux in these TODs, which could either be explained as
a relative calibration error for the two arrays, or as a rapid increase in brightness during the 8
minutes it takes the sky to drift from one array to the next. To investigate this we split each array
into four sub-sets in elevation and measured the flux at the event's position in each.
This resulted in a total of 8 flux measurements taken at roughly 1 minute
intervals. This light curve is shown in Figure~\ref{fig:e1-lightcurve}, and shows a rapid and steady
rise in flux from 300 mJy to a peak of 1100 mJy over the course of 8 minutes, followed by what might be
the start of a slower decay. At this point it was brighter than all but 51 of the 19\,600 point sources
seen by ACT at $>5\sigma$ at f150. We did not detect the event in polarization, from which we place
the 98\% upper limit of polarized flux density of 77 mJy at f090 and 57 mJy at f150.

To rule out calibration errors we measured the light curve for two bright
point sources that were also visible in these TODs. These were flat to within 10\% over this period.
The subsequent observation of the event by two arrays (and their subsets) also rules out terrestrial contamination.
An object fixed to the celestial sphere transits one array in roughly 3 minutes. Because of their position in the focal plane, PA5 and PA6 view a point on the celestial sphere 8 minutes apart. The event was present at the same celestial location for the two TODs, which rules out anything moving quickly, such as an airplane, which would cross one array in roughly a second. Similarly, a source orbiting the Earth would transit an array in roughly 15 seconds.

Since we did not observe this region of sky again in 2019, we cannot say how long the event lasted.

%
%

After the event was identified, we measured the flux at the event's location in maps based on the full
ACT data set from 2008--2019 (of which data from 2016--2019 hit this region) to determine the baseline
flux level. No significant flux was detected and there was nothing out of the ordinary at this position
until the event occurred. We used this to set lower limits on how many times brighter
than its average level the source got at its peak: at 98\% (50\%) confidence the source got at least 36 (88) times
brighter at f090 and 181 (719) times brighter at f150. During the flare the source had a spectral
index of $\alpha=1.5\pm0.2$ as measured between f090 and f150. This value includes a $1\sigma$ calibration
uncertainty of 5\% particular to this analysis. We see no evidence of the spectral index changing
during the flare.

The properties of Event~1 are summarized in the first column of Table~\ref{tab:overview}.

\subsubsection{Possible counterparts}

Event 1 is 3.5 arcsec away from the high proper-motion M star 2MASS J18151564-4927472 at a distance of 62 pc \citep{gaia:2020}.
The star has J2000 coordinates RA=273.8152$^\circ$
Dec=-49.4631$^\circ$ and spectral type M3V.
The star has TESS observations with a $0.4$ day periodicity, which suggests that it is a rapid rotator.
 \citet{messina/etal:2017} notes it as a single-line spectroscopic binary.

The star 2MASS J18151564-4927472 is a member of the $\beta$ Pictoris moving group or possibly the Argus association \citep{moor/etal:2013,2017A&A...607A...3M}.  These associations contain young stars: recent age estimates are in the range 16--28 Myr for the $\beta$ Pic moving group \citep{2016MNRAS.455.3345B,2016A&A...596A..29M,2020A&A...642A.179M} and 40--50 Myr for the Argus association \citep{2019ApJ...870...27Z}.

The association with 2MASS J18151564-4927472 is much closer than one would
expect by chance, given the local density of such stars. According to \citet{gaia:2020}, there are only 1568 stars
of its brightness (g-mag 11.72) or brighter within $2\dg$ of this position.
At this areal density, $n = 9.63\times 10^{-6}$ arcsec$^{-2}$, the probability
of having a star as bright as this within $r = 3.5$ arcsec is only $P = 1 -
\exp(-n\pi r^2) = 3.7\times 10^{-4}$. It therefore seems unlikely that this is
a chance association.

Event 1 is also associated with the ROSAT X-ray source 2RXS J200759.4+160959 \citep{2rxs},
which has a separation of 8.2 arcsec from the event, consistent with the ROSAT position
accuracy of 6 arcsec. The local density of ROSAT X-ray sources of this
brightness (18.83 counts/s) is $1.4 \times 10^{-8} \text{arcesc}^{-2}$ within a
distance of $10^\circ$, leading to a random association probability of just
$2.9 \times 10^{-6}$. The X-ray source is therefore likely to be assicated
with Event 1 and 2MASS J18151564-4927472.

At a distance of 62 pc, the characteristic luminosity (assumed isotropic) of this transient event would be $\nu L_\nu = 2.61 \pm 0.16 \times 10^{22}$ W in f090 and $\nu L_\nu = 7.59 \pm 0.42 \times 10^{22}$ W in f150.
This is roughly a million times brighter than a bright solar flare (e.g. $3.5\times 10^{16}$ W at
212 GHz for an X5.6 class flare \citep{Kaufmann_2003}), but is comparable to giant flares observed
in young stars (see section~\ref{sec:discussion}).

\subsection {Event 2: \eventB{}}

Event 2 was observed at
RA$=105.1584^\circ \pm 0.00085^\circ$ and Dec$=-11.2434^\circ \pm 0.00083^\circ$ in one TOD each of PA4, PA5 and PA6 during the period 02:30--02:38 UTC on 2019-12-15. In that interval,
we observed a flux consistent with being constant ($p = 0.51$, though with large error bars)
at 143 $\pm$ 13.0 mJy at f090 and 304 $\pm$ 18 mJy at f150. The array PA4 is also sensitive
to the f220 frequency band, but the f220 data did not pass our quality cuts for this TOD.
For comparison the mean sky flux at this location across our full data set is consistent with zero,
-0.7/7.7/-5.2 $\pm$ 5.8/6.5/13.2 mJy at f090/f150/f220.
This represents a brightening of a factor of at least 35 (39) at f090 and
11 (14) at f150 at a confidence of 98\% (50\%). We do not detect the event in polarization,
resulting in a 98\% upper limit of 74/63 mJy at f090/f150.

These observations are preceded by an 8 day gap in coverage of this part of the sky, and
this area of the sky was not observed again in the ACT data we have analyzed so far. We can therefore only
limit the rise and fall times for the light curve to be $<$ 8 days and $\gg 10$ min respectively.
The galactic coordinates are $l=-138.01^\circ$ and $b=-3.09^{\circ}$, so the event is close to the galactic
plane. Between f090 and f150, $\alpha=1.8\pm0.2$.

Event 2 is summarized in column~2 in Table~\ref{tab:overview}.

\subsubsection{Possible counterparts}

The two closest sources in the SIMBAD data base \citep{simbad/2000} are
the star HD 52385 (spectral type K0/III) \citep{cannon/pickering:1993,1999MSS...C05....0H}, which is 10.7 arcsec ($2.9\sigma$) away from the position of Event 2, and ROSAT source 2RXS J070037.4-111435\footnote{
	$5.2 \times 10^{-6}$ probability of seeing an X-ray source as bright as its
	22.68 counts/s as close as this by chance.} \citep{2rxs} which is 7.9 arcsec away.

The star's J2000 coordinates are RA=105.1567$^\circ$ Dec=$-11.2459^\circ$ and its distance is 403 pc \citep{gaia:2020}.  It is in the star forming region Canis Major R1.
In a ROSAT study of the young stellar population of this star forming region, the star and X-ray source are also associated by \citet{2009A&A...506..711G}, who from color-magnitude and isochrone fitting determined the object's mass and age as $>5$ $M_\odot$ and $<1$ Myr, although they remained uncertain about the age.

{\sl Gaia} lists 161 stars at least as bright as HD 52385's 8.11 g-mag within $4\dg$ of Event 2, for a local density
of $n=0.247\times 10^{-6}$ arcsec$^{-2}$. The probability for a star as bright as this being
as close as 10.7 arcsec by chance is only $P=8.9\times 10^{-5}$.

At a distance of 403 pc,
the characteristic luminosity (again assuming isotropy) of this event would be
$\nu L_\nu = 27 \pm 2 \times 10^{22}$ W at f090 and
$\nu L_\nu = 89 \pm 5 \times 10^{22}$ W at f150, making it more than ten times as luminous as Event 1.

\begin{figure*}[th]
\begin{center}
\includegraphics[width=0.9\textwidth]{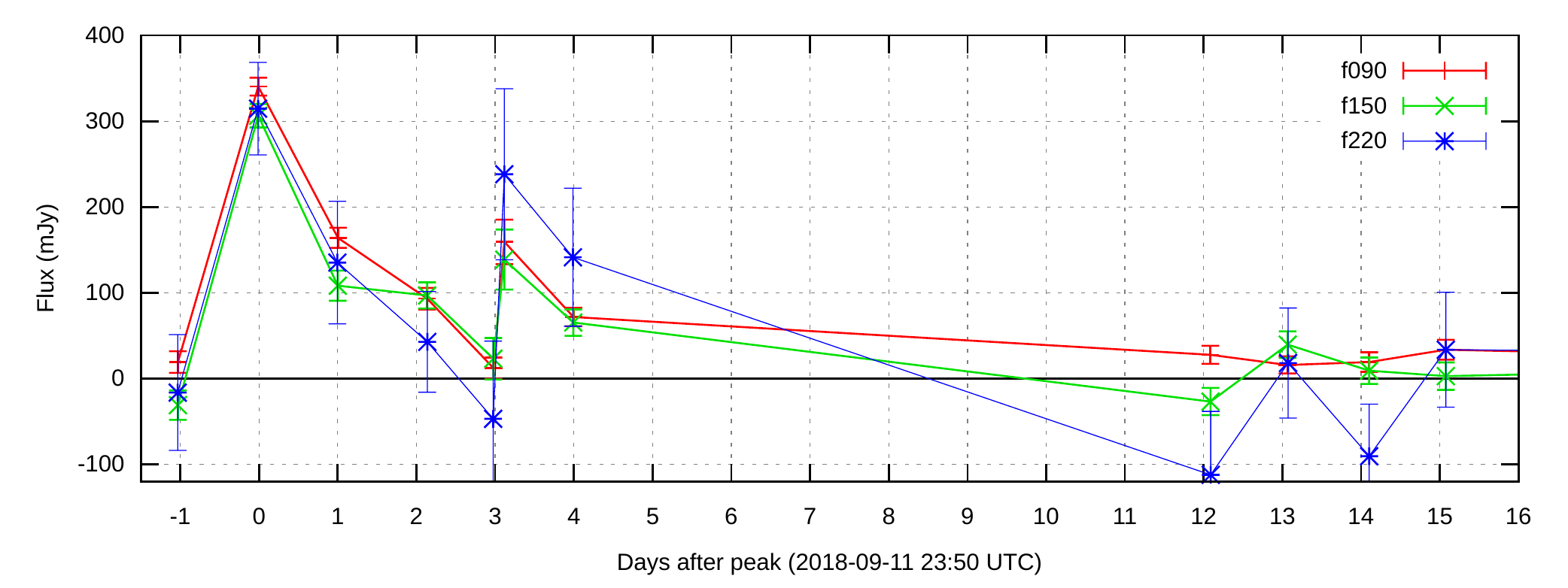}
\caption{The light curve for Event~3 \eventC{}. The event starts with a sudden rise from
below our detection
limit to about 300 mJy in all three of ACT's frequency bands. It then decays roughly linearly to
nearly undetectable over the course of 3 days, after which there is a new jump up to about 150 mJy
followed by a new decay. There are large gaps in our coverage of these coordinates after this, but
by day 12-15 the source is undetectable in f150 and f220 and barely detectable in f090. We do not
detect it after this. In sky maps built from all our data (including the event itself) we detect an object
at this location only in f090, at $3.5 \sigma$.}
\label{fig:event_3_lc}
\end{center}
\end{figure*}

\subsection {Event 3: \eventC{}}

Event 3 was observed at RA$=301.9952^\circ \pm 0.00087^\circ$ and Dec$=16.1652^\circ \pm 0.00083^\circ$
with an (observed) peak on 2018-09-11 23:36, which is also our first detection of a signal at this
location. Unlike the other events, we have extensive, if somewhat irregular,
coverage of this area both before and after the peak, resulting in the light curve shown in
Figure~\ref{fig:event_3_lc}. The source appears from one day to the next with a flux of
340/306/315 $\pm$ 10/14/54 mJy at f090/f150/f220, and then falls
gradually to near non-detection over the course of 3 days, followed by a new sudden rise and
another slow decay. By day 12-15 the source is undetectable in f150 and f220 and barely detectable in f090.
We do not detect the event in polarization, resulting in a 98\% upper limit of
57/95/58 mJy at f090/f150/f220.

When averaged over all our data, we find a mean flux of 10.3/3.6/0.4 $\pm$ 2.9/2.2/4.5 mJy, corresponding
to a $3.6 \sigma$ detection at f090. However, it's important to note this data includes the event itself.
It is possible that the weak detection at f090 in the average data is simply a diluted version of the transient itself,
with no flux being present outside this event.
We will perform a more careful measurement
of the multi-season average flux that excludes the event itself in a future paper describing a systematic search
for ACT transients. If these mean fluxes are representative of the source's behavior outside the
event, then the source got at least 21/39/27 (33/92/98) times brighter in f090/f150/f220 at 98\% (50\%)
confidence.

The galactic coordinates are $l=56.3^\circ$ and $b=-8.8^{\circ}$, again near the galactic plane.
The spectral index at the peak is $-0.25\pm0.17$ between f090 and f150, $-0.23\pm0.44$ between
f090 and f220, and $0.12\pm0.48$ between f150 and f220. We see no evidence of the spectral index changing
during the event.

The properties of Event 3 are summarized in the third column of Table~\ref{tab:overview}.

\subsubsection {Possible counterparts}

The star HD 191179 \citep{cannon/pickering:1993} is within 6.4 arcsec ($1.1\sigma$) with the position of event 3. It is a spectroscopic binary \citep{1998MNRAS.295..257O,2009Obs...129..317G} with J2000 coordinates RA = 301.9968$^\circ$, Dec = 16.1662$^\circ$ and a distance of 219 pc. It is coincident with WISE J200759.23+160958.1
\citep{wright/etal:2010} and X-ray source 1RXS J200759.3+160955\footnote{
	Separation 9.4 arcsec; $2.6 \times 10^{-7}$ probability of seeing an X-ray source as bright as its
	349 counts/s as close as this by chance.} \citep{2000ApJS..131..335R,2012AcA....62...67K,2015A&A...575A..42G}. The X-ray source is identified with the stellar system.
SIMBAD lists this object as a G5 star, but \citet{1998MNRAS.295..257O} and \citet{2002A&A...384..491C} fit the binary system with models of K0IV + G2V (subgiant/dwarf) and K0IV + G2IV (subgiant/subgiant), respectively.   In particular, \citet{2002A&A...384..491C} categorize the G star as a fast rotating solar-type star, which indicates that it is a young object, in a class of stars just before or just after arrival on the main sequence.  The system appears in the catalog of \citet{2008MNRAS.389.1722E} of chromospherically active binary stars.

Although Event 3 has some of the characteristics of a blazar, e.g. a flat spectrum and evolution on day-to-week time-scales, we could not identify a potential candidate.

{\sl Gaia} lists 140 stars at least as bright as HD 191179's 7.96 g-mag within $4\dg$ of Event 3,
for a local density of $n=0.215\times 10^{-6}$ arcsec$^{-2}$. The probability
for a star this bright happening to be within 6.4 arcsec by chance is $P=2.8\times 10^{-5}$. If the
association with HD 191179 is correct, then the peak flux corresponded to a
characteristic luminosity $\nu L_\nu$ of $19.1/26.4/39.7 \pm 0.6/1.2/6.8 \times
10^{22}$ W in f090/f150/f220, making it intermediate between Event 1 and 2.

\section{Discussion}
\label{sec:discussion}

\begin{figure}
	\centering
	\includegraphics[width=\columnwidth]{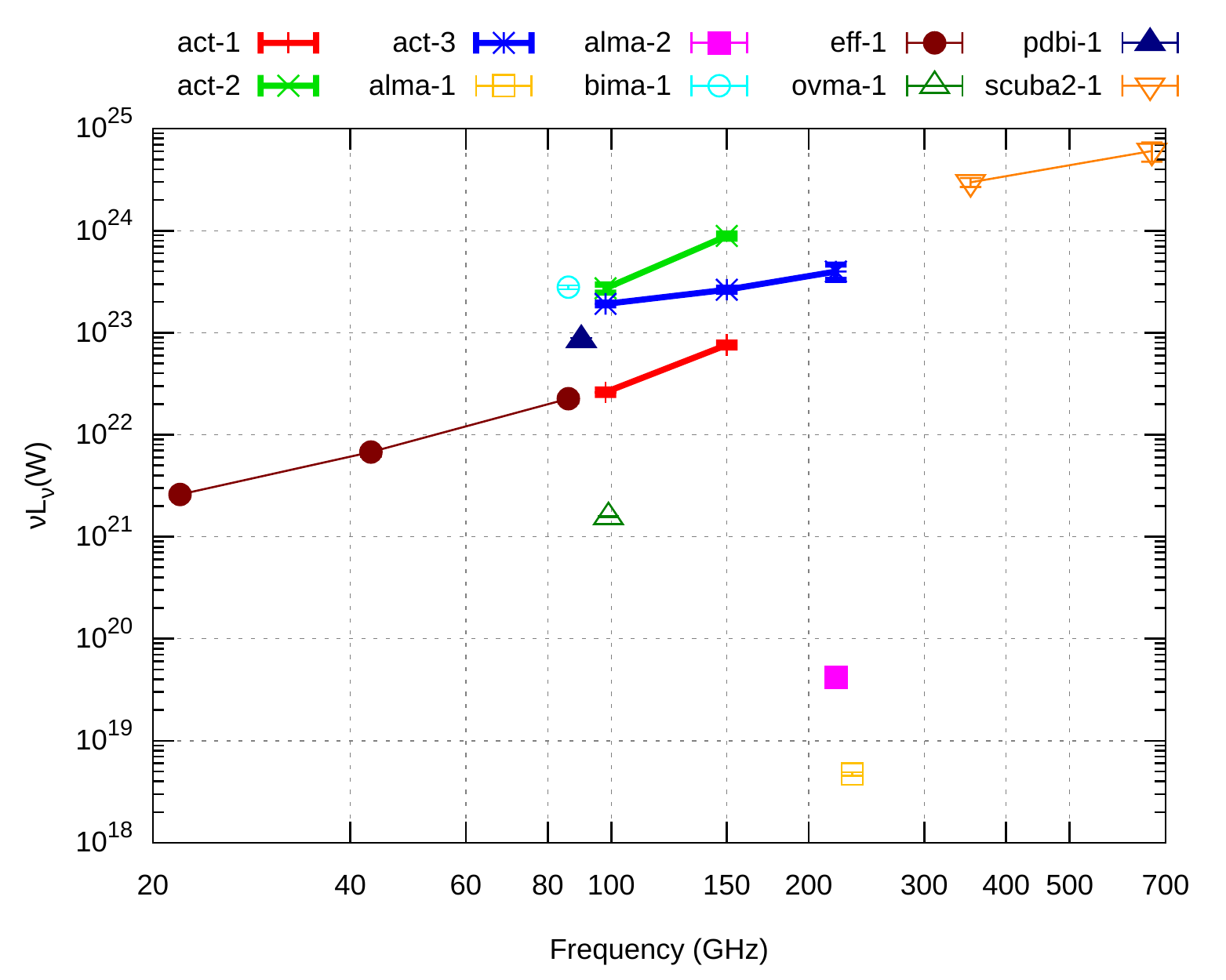}
	\caption{Comparison of the characteristic luminosity with other bright mm-wave star
	flares from the literature. \dfn{act-1/2/3}: Event 1/2/3 from this paper.
	\dfn{alma-1/2}: Proxima Centauri \citep{macgregor/etal:2018} and
	AU Mic \citep{macgregor2020properties} flares measured with ALMA.
	\dfn{bima-1}: Flaring of GMR-A in the Orion Nebula \citep{Bower_2003}.
	\dfn{eff-1, pdbi-1}: V773 Tau by \citet{2009ASPC..402..400U}
	and \citet{Massi_2006} respectively. \dfn{ovma-1}: $\upsigma$ Gem
	\citep{Brown_2006}. \dfn{scuba2-1}: JW 566 \citep{Mairs_2019}.}
	\label{fig:flare_comparison}
\end{figure}

The spatial coincidence of these three sources with stars is notable. Assuming the stellar associations are correct, we can compare them to other mm-wave stellar transient events as in Figure~\ref{fig:flare_comparison}. There is a wide variety of behavior, and our events are comparably luminous to some of the previous observations.  Of those previous events, three are in T Tauri stars:
V773 Tau \citep{2009ASPC..402..400U}, JW 566 \citep{Mairs_2019}, and GMR-A \citep{Bower_2003}.
Three are in binary systems: V773 Tau and JW 566 are binary T Tauri systems, and $\sigma$ Gem is an evolved giant K star in an RS Canum Venaticorum variable binary. Sources like these can exhibit strong
variability in both flux and spectral index during and after a flare \citep{Bower_2003}. The time-scale of these events varies greatly. Some vary on time-scales
of days, while others rise and fall in as little as a minute. For example
Proxima Centauri, an M-type star, increased in brightness by 1000 with a duration of about half a minute at 233 GHz (including both rise and fall), and with a spectral index of $\alpha=-1.77\pm0.45$ \citet{macgregor/etal:2018}.

Likely all of our stars are young.  One star (2MASS J18151564-4927472) is on the main sequence, but is a candidate member of young stellar associations and a fast rotator. The other two are much too luminous to be on the main sequence.
HD 52385 is a young stellar object in a star forming region.  HD  191179 is a fast rotator and young object.
Two of our stars (2MASS J18151564-4927472 and HD 191179) are known binaries. All the stars are associated
with strong X-ray emission.

A main mechanism for stellar flares is magnetic reconnection in coronal loops on the surface of the star.  These can happen in loop collisions on single stars, but are enhanced in young stellar objects by interactions with the protoplanetary disc and in binary star systems by interactions between the coronas of the two stars \citep{Massi_2006}.

We expect to detect more flaring stars as well as extragalactic transients in a future systematic search for transients in the ACT data. This will permit an assessment of the associated event rates. The large area mm-wave coverage of ACT and the upcoming Simons Observatory \citep{SO:2019} and CMB-S4 \citep{CMB-S4:2019} will nicely complement the Vera Rubin Observatory's detection of transients in the optical \citep{lsst-science-drivers-2019}.
The lack of time coverage in our detections highlights the need for a regular and frequent cadence to characterize such events well.

By surveying large areas, mm-wave instruments will be able to measure the population statistics of these mm-wave stellar flares.  With a blind search, we will be able to constrain the rates of such flares both inside star forming regions (where many previous works have looked) and outside them in the general stellar population.

\section*{Acknowledgments}
This work was supported by the U.S. National Science Foundation through awards AST-0408698, AST-0965625, and AST-1440226 for the ACT project, as well as awards PHY-0355328, PHY-0855887 and PHY-1214379. Funding was also provided by Princeton University, the University of Pennsylvania, and a Canada Foundation for Innovation (CFI) award to UBC. ACT operates in the Parque Astron\'omico Atacama in northern Chile under the auspices of the Comisi\'on Nacional de Investigaci\'on (CONICYT).
Computations were performed on the Niagara supercomputer at the SciNet HPC Consortium and on the Simons-Popeye cluster of the Flatiron Institute. SciNet is funded by the CFI under the auspices of Compute Canada, the Government of Ontario, the Ontario Research Fund---Research Excellence, and the University of Toronto.
KMH is supported by NSF through AST 1815887.
EC acknowledges support from the STFC Ernest Rutherford Fellowship ST/M004856/2 and STFC Consolidated Grant ST/S00033X/1, and from the European Research Council (ERC) under the European Union’s Horizon 2020 research and innovation programme (Grant agreement No. 849169).
SKC acknowledges support from NSF award AST-2001866.
ZX is supported by the Gordon and Betty Moore Foundation.
NS acknowledges support from NSF grant number AST-1907657.

We gratefully acknowledge the many publicly available software packages that were essential for parts of this analysis. They include
\texttt{healpy}~\citep{Healpix1}, \texttt{HEALPix}~\citep{Healpix2}, and
\texttt{pixell}\footnote{https://github.com/simonsobs/pixell}. This research made use of \texttt{Astropy}\footnote{http://www.astropy.org}, a community-developed core Python package for Astronomy \citep{astropy:2013, astropy:2018}. We also acknowledge use of the \texttt{matplotlib}~\citep{Hunter:2007} package and the Python Image Library for producing plots in this paper.
Lastly, we thank Rachel Osten for especially helpful comments on an earlier draft of this paper. 

\bibliographystyle{act_titles}
\bibliography{act_mm_refs,apj-jour}

\begin{thebibliography}{56}
\expandafter\ifx\csname natexlab\endcsname\relax\def\natexlab#1{#1}\fi

\bibitem[{{Abazajian} et~al.(2019)}]{CMB-S4:2019}
{Abazajian}, K., et~al. 2019,
  \href{https://arxiv.org/abs/1907.04473}{\texttt{arXiv:1907.04473}}, arXiv
  e-prints, arXiv:1907.04473, {CMB-S4 Science Case, Reference Design, and
  Project Plan}

\bibitem[{{Ade} et~al.(2019)}]{SO:2019}
{Ade}, P., et~al. 2019,
  \href{https://arxiv.org/abs/1808.07445}{\texttt{arXiv:1808.07445}}, \jcap,
  2019, 056, {The Simons Observatory: science goals and forecasts}

\bibitem[{{Aiola} et~al.(2020)}]{aiola/etal:2020}
{Aiola}, S., et~al. 2020,
  \href{https://arxiv.org/abs/2007.07288}{\texttt{arXiv:2007.07288}}, arXiv
  e-prints, arXiv:2007.07288, {The Atacama Cosmology Telescope: DR4 Maps and
  Cosmological Parameters}

\bibitem[{{Astropy Collaboration}(2013)}]{astropy:2013}
{Astropy Collaboration}. 2013,
  \href{https://arxiv.org/abs/1307.6212}{\texttt{arXiv:1307.6212}}, \aap, 558,
  A33, {Astropy: A community Python package for astronomy}

\bibitem[{{Batygin} et~al.(2019){Batygin}, {Adams}, {Brown}, \&
  {Becker}}]{p9_hypothesis}
{Batygin}, K., {Adams}, F.~C., {Brown}, M.~E., \& {Becker}, J.~C. 2019,
  \href{https://arxiv.org/abs/1902.10103}{\texttt{arXiv:1902.10103}}, \physrep,
  805, 1, {The planet nine hypothesis}

\bibitem[{{Binks} \& {Jeffries}(2016)}]{2016MNRAS.455.3345B}
{Binks}, A.~S. \& {Jeffries}, R.~D. 2016,
  \href{https://arxiv.org/abs/1510.06987}{\texttt{arXiv:1510.06987}}, \mnras,
  455, 3345, {Spectroscopic confirmation of M-dwarf candidate members of the
  Beta Pictoris and AB Doradus Moving Groups}

\bibitem[{{Boller} et~al.(2016){Boller}, {Freyberg}, {Tr{\"u}mper}, {Haberl},
  {Voges}, \& {Nandra}}]{2rxs}
{Boller}, T., {Freyberg}, M.~J., {Tr{\"u}mper}, J., {Haberl}, F., {Voges}, W.,
  \& {Nandra}, K. 2016,
  \href{https://arxiv.org/abs/1609.09244}{\texttt{arXiv:1609.09244}}, \aap,
  588, A103, {Second ROSAT all-sky survey (2RXS) source catalogue}

\bibitem[{Bower et~al.(2003)Bower, Plambeck, Bolatto, McCrady, Graham,
  de~Pater, Liu, \& Baganoff}]{Bower_2003}
Bower, G.~C., Plambeck, R.~L., Bolatto, A., McCrady, N., Graham, J.~R.,
  de~Pater, I., Liu, M.~C., \& Baganoff, F.~K. 2003,
  \href{https://arxiv.org/abs/astro-ph/0308277}{\texttt{astro-ph/0308277}},
  \apj, 598, 1140–1150, A Giant Outburst at Millimeter Wavelengths in the
  Orion Nebula

\bibitem[{Brown \& Brown(2006)}]{Brown_2006}
Brown, J.~M. \& Brown, A. 2006, \apj, 638, L37, A Large Millimeter Flare on the
  {RS} {CVn} Binary $\upsigma$ Geminorum

\bibitem[{{Cannon} \& {Pickering}(1993)}]{cannon/pickering:1993}
{Cannon}, A.~J. \& {Pickering}, E.~C. 1993, VizieR Online Data Catalog,
  III/135A, {VizieR Online Data Catalog: Henry Draper Catalogue and Extension
  (Cannon+ 1918-1924; ADC 1989)}

\bibitem[{{Choi} et~al.(2018)}]{choi/etal:2018}
{Choi}, S.~K., et~al. 2018,
  \href{https://arxiv.org/abs/1711.04841}{\texttt{arXiv:1711.04841}}, JLTP,
  193, 267, {Characterization of the Mid-Frequency Arrays for Advanced ACTPol}

\bibitem[{{Choi} et~al.(2020)}]{choi/etal:2020}
---. 2020, \href{https://arxiv.org/abs/2007.07289}{\texttt{arXiv:2007.07289}},
  arXiv e-prints, arXiv:2007.07289, {The Atacama Cosmology Telescope: A
  Measurement of the Cosmic Microwave Background Power Spectra at 98 and 150
  GHz}

\bibitem[{{Crowley} et~al.(2018)}]{crowley/etal:2018}
{Crowley}, K.~T., et~al. 2018,
  \href{https://arxiv.org/abs/1807.07496}{\texttt{arXiv:1807.07496}}, JLTP,
  193, 328, {Advanced ACTPol TES Device Parameters and Noise Performance in
  Fielded Arrays}

\bibitem[{{Cutispoto} et~al.(2002){Cutispoto}, {Pastori}, {Pasquini}, {de
  Medeiros}, {Tagliaferri}, \& {Andersen}}]{2002A&A...384..491C}
{Cutispoto}, G., {Pastori}, L., {Pasquini}, L., {de Medeiros}, J.~R.,
  {Tagliaferri}, G., \& {Andersen}, J. 2002, \aap, 384, 491, {Fast-rotating
  nearby solar-type stars, Li abundances and X-ray luminosities. I. Spectral
  classification, v sin i, Li abundances and X-ray luminosities}

\bibitem[{{De Bernardis} et~al.(2016)}]{debernardis/etal:2016}
{De Bernardis}, F., et~al. 2016,
  \href{https://arxiv.org/abs/1607.02120}{\texttt{arXiv:1607.02120}}, in
  Society of Photo-Optical Instrumentation Engineers (SPIE) Conference Series,
  Vol. 9910, Observatory Operations: Strategies, Processes, and Systems VI, ed.
  A.~B. {Peck}, R.~L. {Seaman}, \& C.~R. {Benn},  991014

\bibitem[{{de Ugarte Postigo} et~al.(2012)}]{2012A&A...538A..44D}
{de Ugarte Postigo}, A., et~al. 2012,
  \href{https://arxiv.org/abs/1108.1797}{\texttt{arXiv:1108.1797}}, \aap, 538,
  A44, {Pre-ALMA observations of GRBs in the mm/submm range}

\bibitem[{{Eker} et~al.(2008)}]{2008MNRAS.389.1722E}
{Eker}, Z., et~al. 2008,
  \href{https://arxiv.org/abs/0805.4517}{\texttt{arXiv:0805.4517}}, \mnras,
  389, 1722, {A catalogue of chromospherically active binary stars (third
  edition)}

\bibitem[{{Gaia Collaboration}(2020)}]{gaia:2020}
{Gaia Collaboration}. 2020,
  \href{https://arxiv.org/abs/2012.01533}{\texttt{arXiv:2012.01533}}, Gaia
  Early Data Release 3: Summary of the contents and survey properties

\bibitem[{{G{\'o}rski} et~al.(2005){G{\'o}rski}, {Hivon}, {Banday}, {Wandelt},
  {Hansen}, {Reinecke}, \& {Bartelmann}}]{Healpix2}
{G{\'o}rski}, K.~M., {Hivon}, E., {Banday}, A.~J., {Wandelt}, B.~D., {Hansen},
  F.~K., {Reinecke}, M., \& {Bartelmann}, M. 2005,
  \href{https://arxiv.org/abs/arXiv:astro-ph/0409513}{\texttt{arXiv:astro-ph/0409513}},
  \apj, 622, 759, {HEALPix: A Framework for High-Resolution Discretization and
  Fast Analysis of Data Distributed on the Sphere}

\bibitem[{{Gregorio-Hetem} et~al.(2009){Gregorio-Hetem}, {Montmerle},
  {Rodrigues}, {Marciotto}, {Preibisch}, \& {Zinnecker}}]{2009A&A...506..711G}
{Gregorio-Hetem}, J., {Montmerle}, T., {Rodrigues}, C.~V., {Marciotto}, E.,
  {Preibisch}, T., \& {Zinnecker}, H. 2009,
  \href{https://arxiv.org/abs/0909.2888}{\texttt{arXiv:0909.2888}}, \aap, 506,
  711, {Star formation history of Canis Major R1. I. Wide-Field X-ray study of
  the young stellar population}

\bibitem[{{Greiner} \& {Richter}(2015)}]{2015A&A...575A..42G}
{Greiner}, J. \& {Richter}, G.~A. 2015,
  \href{https://arxiv.org/abs/1408.5529}{\texttt{arXiv:1408.5529}}, \aap, 575,
  A42, {Optical counterparts of ROSAT X-ray sources in two selected fields at
  low vs. high Galactic latitudes}

\bibitem[{{Griffin}(2009)}]{2009Obs...129..317G}
{Griffin}, R.~F. 2009, Obs, 129, 317, {Spectroscopic binary orbits from
  photoelectric radial velocities. Paper 209: Twenty short-period
  active-chromosphere stars}

\bibitem[{Hasselfield(2013)}]{matthew-thesis}
Hasselfield, M. 2013, Ph.D. thesis, University of British Columbia

\bibitem[{{Henderson} et~al.(2016)}]{henderson/etal:2016}
{Henderson}, S.~W., et~al. 2016,
  \href{https://arxiv.org/abs/1510.02809}{\texttt{arXiv:1510.02809}}, JLTP,
  184, 772, {Advanced ACTPol Cryogenic Detector Arrays and Readout}

\bibitem[{{Ho} et~al.(2019)}]{ho/etal:2019}
{Ho}, A. Y.~Q., et~al. 2019,
  \href{https://arxiv.org/abs/1810.10880}{\texttt{arXiv:1810.10880}}, \apj,
  871, 73, {AT2018cow: A Luminous Millimeter Transient}

\bibitem[{{Houk} \& {Swift}(1999)}]{1999MSS...C05....0H}
{Houk}, N. \& {Swift}, C. 1999, Michigan Spectral Survey, 5, 0, {Michigan
  catalogue of two-dimensional spectral types for the HD Stars, Vol. 5}

\bibitem[{Hunter(2007)}]{Hunter:2007}
Hunter, J.~D. 2007, Computing in Science \& Engineering, 9, 90, Matplotlib: A
  2D graphics environment

\bibitem[{{Ivezi{\'c}} et~al.(2019)}]{lsst-science-drivers-2019}
{Ivezi{\'c}}, {\v Z}., et~al. 2019,
  \href{https://arxiv.org/abs/0805.2366}{\texttt{arXiv:0805.2366}}, \apj, 873,
  111, {LSST: From Science Drivers to Reference Design and Anticipated Data
  Products}

\bibitem[{Kaufmann et~al.(2002)}]{Kaufmann_2003}
Kaufmann, P., et~al. 2002, \apj, 574, 1059, Solar Submillimeter and Gamma-Ray
  Burst Emission

\bibitem[{{Kiraga}(2012)}]{2012AcA....62...67K}
{Kiraga}, M. 2012,
  \href{https://arxiv.org/abs/1204.3825}{\texttt{arXiv:1204.3825}}, \actaa, 62,
  67, {ASAS Photometry of ROSAT Sources. I. Periodic Variable Stars Coincident
  with Bright Sources from the ROSAT All Sky Survey}

\bibitem[{{Kuno} et~al.(2004){Kuno}, {Sato}, {Nakanishi}, {Yamauchi}, {Nakai},
  \& {Kawai}}]{2004PASJ...56L...1K}
{Kuno}, N., {Sato}, N., {Nakanishi}, H., {Yamauchi}, A., {Nakai}, N., \&
  {Kawai}, N. 2004,
  \href{https://arxiv.org/abs/astro-ph/0401258}{\texttt{astro-ph/0401258}},
  \pasj, 56, L1, {Radio Observations of the Afterglow of GRB 030329}

\bibitem[{{Laskar} et~al.(2019)}]{laskar-grb-2019}
{Laskar}, T., et~al. 2019,
  \href{https://arxiv.org/abs/1904.07261}{\texttt{arXiv:1904.07261}}, \apjl,
  878, L26, {ALMA Detection of a Linearly Polarized Reverse Shock in GRB
  190114C}

\bibitem[{{Lei} et~al.(2016){Lei}, {Yuan}, {Zhang}, \& {Wang}}]{Lei_2016}
{Lei}, W.-H., {Yuan}, Q., {Zhang}, B., \& {Wang}, D. 2016,
  \href{https://arxiv.org/abs/1511.01206}{\texttt{arXiv:1511.01206}}, \apj,
  816, 20, {IGR J12580+0134: The First Tidal Disruption Event with an Off-beam
  Relativistic Jet}

\bibitem[{MacGregor et~al.(2020)MacGregor, Osten, \&
  Hughes}]{macgregor2020properties}
MacGregor, M.~A., Osten, R.~A., \& Hughes, A.~M. 2020,
  \href{https://arxiv.org/abs/2001.10546}{\texttt{arXiv:2001.10546}},
  Properties of M Dwarf Flares at Millimeter Wavelengths

\bibitem[{MacGregor et~al.(2018)MacGregor, Weinberger, Wilner, Kowalski, \&
  Cranmer}]{macgregor/etal:2018}
MacGregor, M.~A., Weinberger, A.~J., Wilner, D.~J., Kowalski, A.~F., \&
  Cranmer, S.~R. 2018,
  \href{https://arxiv.org/abs/1802.08257}{\texttt{arXiv:1802.08257}}, \apj,
  855, L2, Detection of a Millimeter Flare from Proxima Centauri

\bibitem[{Mairs et~al.(2019)}]{Mairs_2019}
Mairs, S., et~al. 2019,
  \href{https://arxiv.org/abs/1812.00016}{\texttt{arXiv:1812.00016}}, \apj,
  871, 72, The JCMT Transient Survey: An Extraordinary Submillimeter Flare in
  the T Tauri Binary System JW 566

\bibitem[{Massi et~al.(2006)Massi, Forbrich, Menten, Torricelli-Ciamponi,
  Neidhöfer, Leurini, \& Bertoldi}]{Massi_2006}
Massi, M., Forbrich, J., Menten, K.~M., Torricelli-Ciamponi, G., Neidhöfer,
  J., Leurini, S., \& Bertoldi, F. 2006,
  \href{https://arxiv.org/abs/astro-ph/0604124}{\texttt{astro-ph/0604124}},
  \aap, 453, 959–964, Synchrotron emission from the T Tauri binary system
  V773 Tauri A

\bibitem[{{Messina} et~al.(2016)}]{2016A&A...596A..29M}
{Messina}, S., et~al. 2016,
  \href{https://arxiv.org/abs/1607.06634}{\texttt{arXiv:1607.06634}}, \aap,
  596, A29, {The rotation-lithium depletion correlation in the
  {\ensuremath{\beta}} Pictoris association and the LDB age determination}

\bibitem[{{Messina} et~al.(2017{\natexlab{a}})}]{messina/etal:2017}
---. 2017{\natexlab{a}},
  \href{https://arxiv.org/abs/1612.04591}{\texttt{arXiv:1612.04591}}, \aap,
  600, A83, {The {\ensuremath{\beta}} Pictoris association: Catalog of
  photometric rotational periods of low-mass members and candidate members}

\bibitem[{{Messina} et~al.(2017{\natexlab{b}})}]{2017A&A...607A...3M}
---. 2017{\natexlab{b}},
  \href{https://arxiv.org/abs/1707.01682}{\texttt{arXiv:1707.01682}}, \aap,
  607, A3, {The {\ensuremath{\beta}} Pictoris association low-mass members:
  Membership assessment, rotation period distribution, and dependence on
  multiplicity}

\bibitem[{{Miret-Roig} et~al.(2020)}]{2020A&A...642A.179M}
{Miret-Roig}, N., et~al. 2020,
  \href{https://arxiv.org/abs/2007.10997}{\texttt{arXiv:2007.10997}}, \aap,
  642, A179, {Dynamical traceback age of the {\ensuremath{\beta}} Pictoris
  moving group}

\bibitem[{{Mo{\'o}r} et~al.(2013){Mo{\'o}r}, {Szab{\'o}}, {Kiss}, {Kiss},
  {{\'A}brah{\'a}m}, {Szul{\'a}gyi}, {K{\'o}sp{\'a}l}, \&
  {Szalai}}]{moor/etal:2013}
{Mo{\'o}r}, A., {Szab{\'o}}, G.~M., {Kiss}, L.~L., {Kiss}, C.,
  {{\'A}brah{\'a}m}, P., {Szul{\'a}gyi}, J., {K{\'o}sp{\'a}l}, {\'A}., \&
  {Szalai}, T. 2013,
  \href{https://arxiv.org/abs/1309.1669}{\texttt{arXiv:1309.1669}}, \mnras,
  435, 1376, {Unveiling new members in five nearby young moving groups}

\bibitem[{{Naess} et~al.(2020)}]{naess/etal:2020}
{Naess}, S., et~al. 2020,
  \href{https://arxiv.org/abs/2007.07290}{\texttt{arXiv:2007.07290}}, arXiv
  e-prints, arXiv:2007.07290, {The Atacama Cosmology Telescope:
  arcminute-resolution maps of 18,000 square degrees of the microwave sky from
  ACT 2008-2018 data combined with Planck}

\bibitem[{Naess et~al.(early 2021)}]{act-planet9}
Naess, S. et~al. early 2021,
  \href{https://arxiv.org/abs/2102.????}{\texttt{arXiv:2102.????}}, The Atacama
  Cosmology Telescope: A search for Planet 9 [in preparation]

\bibitem[{{Osten} \& {Saar}(1998)}]{1998MNRAS.295..257O}
{Osten}, R.~A. \& {Saar}, S.~H. 1998, \mnras, 295, 257, {Physical properties of
  active stars and stellar systems}

\bibitem[{{Price-Whelan} et~al.(2018)}]{astropy:2018}
{Price-Whelan}, A.~M., et~al. 2018, \aj, 156, 123, {The Astropy Project:
  Building an Open-science Project and Status of the v2.0 Core Package}

\bibitem[{{Rutledge} et~al.(2000){Rutledge}, {Brunner}, {Prince}, \&
  {Lonsdale}}]{2000ApJS..131..335R}
{Rutledge}, R.~E., {Brunner}, R.~J., {Prince}, T.~A., \& {Lonsdale}, C. 2000,
  \href{https://arxiv.org/abs/astro-ph/0004053}{\texttt{astro-ph/0004053}},
  \apjs, 131, 335, {XID: Cross-Association of ROSAT/Bright Source Catalog X-Ray
  Sources with USNO A-2 Optical Point Sources}

\bibitem[{{Rybicki} \& {Lightman}(1986)}]{1986rpa..book.....R}
{Rybicki}, G.~B. \& {Lightman}, A.~P. 1986, {Radiative Processes in
  Astrophysics} (Wiley-VCH)

\bibitem[{{Thornton} et~al.(2016)}]{thornton/etal:2016}
{Thornton}, R.~J., et~al. 2016,
  \href{https://arxiv.org/abs/1605.06569}{\texttt{arXiv:1605.06569}}, \apjs,
  227, 21, {The Atacama Cosmology Telescope: The Polarization-sensitive ACTPol
  Instrument}

\bibitem[{{Umemoto} et~al.(2009){Umemoto}, {Saito}, {Nakanishi}, {Kuno}, \&
  {Tsuboi}}]{2009ASPC..402..400U}
{Umemoto}, T., {Saito}, M., {Nakanishi}, K., {Kuno}, N., \& {Tsuboi}, M. 2009,
  in Astronomical Society of the Pacific Conference Series, Vol. 402,
  Approaching Micro-Arcsecond Resolution with VSOP-2: Astrophysics and
  Technologies, ed. Y.~{Hagiwara}, E.~{Fomalont}, M.~{Tsuboi}, \&
  M.~{Yasuhiro},  400

\bibitem[{{Wenger} et~al.(2000)}]{simbad/2000}
{Wenger}, M., et~al. 2000,
  \href{https://arxiv.org/abs/astro-ph/0002110}{\texttt{astro-ph/0002110}},
  \aaps, 143, 9, {The SIMBAD astronomical database. The CDS reference database
  for astronomical objects}

\bibitem[{{Whitehorn} et~al.(2016)}]{whitehorn/etal:2016}
{Whitehorn}, N., et~al. 2016,
  \href{https://arxiv.org/abs/1604.03507}{\texttt{arXiv:1604.03507}}, \apj,
  830, 143, {Millimeter Transient Point Sources in the SPTpol 100 Square Degree
  Survey}

\bibitem[{{Wright} et~al.(2010)}]{wright/etal:2010}
{Wright}, E.~L., et~al. 2010,
  \href{https://arxiv.org/abs/1008.0031}{\texttt{arXiv:1008.0031}}, \aj, 140,
  1868, {The Wide-field Infrared Survey Explorer (WISE): Mission Description
  and Initial On-orbit Performance}

\bibitem[{{Yuan} et~al.(2016){Yuan}, {Wang}, {Lei}, {Gao}, \&
  {Zhang}}]{2016MNRAS.461.3375Y}
{Yuan}, Q., {Wang}, Q.~D., {Lei}, W.-H., {Gao}, H., \& {Zhang}, B. 2016,
  \href{https://arxiv.org/abs/1606.06830}{\texttt{arXiv:1606.06830}}, \mnras,
  461, 3375, {Catching jetted tidal disruption events early in millimetre}

\bibitem[{Zonca et~al.(2019)Zonca, Singer, Lenz, Reinecke, Rosset, Hivon, \&
  Gorski}]{Healpix1}
Zonca, A., Singer, L., Lenz, D., Reinecke, M., Rosset, C., Hivon, E., \&
  Gorski, K. 2019, JOSS, 4, 1298, healpy: equal area pixelization and spherical
  harmonics transforms for data on the sphere in Python

\bibitem[{{Zuckerman}(2019)}]{2019ApJ...870...27Z}
{Zuckerman}, B. 2019,
  \href{https://arxiv.org/abs/1811.01508}{\texttt{arXiv:1811.01508}}, \apj,
  870, 27, {The Nearby, Young, Argus Association: Membership, Age, and Dusty
  Debris Disks}

\end{thebibliography}

\end{document}